# Soft Realization: a Bio-inspired Implementation Paradigm

*H.R. Mahdiani[1,2], M. N. Bojnordi[3], and S. M. Fakhraie[4]*
[1] Computer Science and Engineering Department, Shahid Beheshti University, Tehran, IRAN
[2] Institute for Cognitive and Brain Science, Shahid Beheshti University, Tehran, IRAN
[3] School of Computing, University of Utah, Salt Lake City, USA
[4] Electrical and Computer Engineering Department, University of Tehran, Tehran, IRAN

**Abstract**
Researchers traditionally solve the computational problems through rigorous and deterministic algorithms called as *Hard Computing*. Except for some rare analog implementations, these precise algorithms have widely been realized using digital technology as an inherently reliable and accurate implementation platform, either in hardware or software forms. This rigid form of implementation which we refer as *Hard Realization* relies on strict algorithmic accuracy constraints dictated to digital design engineers. Regardless of the type of application, hard realization admits paying as much as necessary implementation costs to preserve computation precision and determinism throughout all the design and implementation steps. Despite all prior accomplishments in hard realization, this conventional paradigm has encountered serious challenges with today's emerging applications and implementation technologies. Unlike traditional hard computing, the emerging soft and bio-inspired algorithms do not rely on fully precise and deterministic computation, thereby not fully satisfied by old-fashion implementations. Moreover, the incoming nanotechnologies face increasing reliability issues that prevent them from being efficiently exploited in hard realization of applications. This article examines *Soft Realization*, a novel bio-inspired approach to design and implementation of an important category of applications noticing the internal brain structure. The proposed paradigm mitigates major weaknesses of hard realization by (1) alleviating incompatibilities with today's soft and bio-inspired algorithms such as artificial neural networks, fuzzy systems, and human sense signal processing applications, and (2) resolving the destructive inconsistency with unreliable nanotechnologies. Our experimental results on a set of well-known soft applications implemented using the proposed soft realization paradigm in both reliable and unreliable technologies indicate that significant energy, delay, and area savings can be obtained compared to the conventional implementation.

## 1- Hard Realization: The Most Popular Implementation Paradigm Ever
Realizing an efficient system requires (1) developing an appropriate algorithm for solving the target problem, and (2) implementing the algorithm using an underlying technology that satisfies user objectives such as speed, reliability, and energy-efficiency. Conventional system designers traditionally solve problems by exploring deterministic, exact, and explicit algorithms developed using analytical models and precise datasets. This well-defined and rigorous problem solving paradigm— a.k.a., *Hard Computing*—requires the implementation platform to guarantee full accuracy and all of the reliability requirements[1]. This strict algorithmic demand has been granted by digital technology since decades ago, thereby enabling today's digital world of sophisticated applications across various disciplines, from science and engineering to economics and transportation. Digital technology has been mostly preferred over other alternatives such as analog implementation as the most important rival, due to providing high computation reliability and precision, noise immunity, and repeatable and predictable behavior. Moreover, traditional digital CMOS technologies are employed to build operational error-free compute substrates with always operational transistors and interconnects because of their high integration density ($10^8$ to $10^9$ transistors/cm$^2$) and low defect rate ($10^{-7}$ to$10^{-9}$ failures/part)[2]. These features have turned the digital technology either in hardware or software forms into the most favored implementation choice for most complex and computationally intensive hard computing algorithms which demand for bit-level precision guarantee and fully deterministic decision making capabilities.

Hard realization was born decades ago and matured over time based on the strong affinity between basic hard computing requirements and capabilities of digital technology. As a result of this maturity, it is now possible to either directly compile a complex precise algorithm on a digital ASIC silicon chip integrating billions of transistors that work in parallel, or realize it as a verified software package comprising millions of machine code lines being executed on a Von Neumann architecture. Although these two approaches lead to different physical trade-offs, they are well-known hard realized instances of deterministic hard computing applications using predictable, deterministic, and precise binary logic. For example, designing a 64-bit hardware adder with hard realization leads into paying necessary amounts of area, delay, and power overheads to preserve comprehensive precision for all the result bits regardless of their significance. Similarly, in software implementation, all parts of the computer (including hardware units such as ALU, registers, and memory, as well as the operating system and compiler) are developed so to preserve complete bit-level accuracy of the software.

Due to pervasive use of hard realization nearly in all existing digital systems, introducing any change to this paradigm may result in a significant impact on various aspects of today's digital world. On the other hand, current hard realization techniques are supported by numerous ready-to-use CAD tools, compilers, and debuggers that make it very hard to tweak all these tools to follow any fundamental change in hard realization basics. Today however, there are some inevitable evidences that enforce the digital world to accept some vital changes in conventional hard realization. The rest of this paper is organized as follows. We investigate the key difficulties of hard realization with the emerging algorithms and nanotechnologies in Section 2. In Section 3, we introduce a new broad and significant category of applications named as *Imprecision Tolerant* which inherently can tolerate some imprecision levels and thus are not efficiently implemented using hard realization. We then propose a new implementation paradigm named as *Soft Realization* specifically for implementing imprecision tolerant applications in Section 4. To assess the efficiency of the proposed paradigm, two major technical instances of soft realization methodologies are studied in Section 5. The conclusion is the last section.

**2- Key Difficulties in Hard Realization with Emerging Applications and Technologies**
Table I summarizes the benefits and shortcomings of hard realization when it is exploited to implement a traditional or emerging application on either a traditional or emerging implementation technology. Based on previous explanations, the top-left cell of the table implies total compatibility of hard realization with any precise technology. Although hard realization has been successfully employed to realize many digital VLSI systems over the past decades, the other cells of the table indicate that exploiting it to realize emerging applications either on existing or evolving implementation technologies has caused significant pitfalls. The main reason is that hard realization foundations were initially based on old-style algorithmic and technological requirements that inherently differ from those of the emerging algorithms and technologies. The advent of many soft and nondeterministic problem solving methodologies as well as the appearance of significant reliability issues in nano-level implementation technologies urge an inevitable paradigm shift from long-established boundaries of the conventional hard realization to gain better implementations compared to old-fashion realization paradigm.

**2-1- Hard Realization Inconsistency Towards Soft and Bio-inspired Algorithms**
Despite recent advances in digital technology, emerging applications seem to require substantially more processing power than the state-of-the-art computer systems. Instances of the emerging computationally intensive applications are advanced data classification, pattern recognition, feature extraction, reasoning applications, cognitive science, brain-computer interface, natural language processing, and human-machine interface, as well as an extensive range of signal, voice, and video processing. On the other hand, a simple investigation of many existing nature-made creatures show that the biological organisms and societies often outperform their artificial rivals in numerous real-world applications[3]. Based on this fact that has been noticed by many people, there are increasing numbers of researchers who believe that the best (and perhaps the only) way to achieve the similar performance of the biological systems is to discover their inherent problem solving structure and then develop an artificial rival using bio-inspiration[3]. Bio-inspiration is not a new methodology; however, it has reemerged due to recent advances in capturing internal structure of the biological systems and the appearance of enabler technologies. Evidence of this reemergence is the increasing number of imprecise, lossy, and approximate algorithms that basically contradicted with the traditional hard computing solutions.

Table I- Advantages and shortcomings of *Hard Realization* for implementing traditional and emerging algorithms using existing or future implementation technologies.

|  | **Traditional Algorithms** (Hard & Rigorous) | **Emerging Algorithms** (Soft & Bio-inspired) |
|---|---|---|
| **Traditional Technologies** (Deterministic & Precise) | Hard realization is historically developed based on specifications of hard computing and digital technology. Therefore, it is compatible with precise and deterministic nature of the traditional algorithms and implementation technologies. | The achieved full accuracy dictated by hard realization is more than needed by soft algorithms. This simply imposes more extra precision cost without any further performance improvements. |
| **Emerging Nano-Technologies** (Nondeterministic & Imprecise) | Due to inherent unreliability of the emerging technologies, it is mandatory to exploit fault-tolerant design techniques and pay high overheads to preserve full accuracy demanded by hard computing applications and dictated by hard realization. However, it also might be sometimes impossible to preserve full accuracy even by paying any overhead cost! | Due to inherent unreliability of the emerging technologies, high overhead fault-tolerant design techniques should be exploited to preserve full accuracy dictated by hard realization, if it is possible of course! But the achieved accuracy is more than needed by soft algorithms and only increases implementation costs without any performance improvement! |

Beside higher performance, another valuable specification of the bio-inspired solutions is that they also inherit some other useful behavioral and structural characteristics from biological systems. The distributed structure and massive parallelism of the bio-inspired solutions, as well as their capability for providing approximate solutions in short-term as well as precise solutions in longer periods, should be treated as some of their other additional advantages over hard computing. Figure 1 demonstrates some well-known instances of biological phenomena and some of their useful common properties. Moreover, as the biological systems do not rely on a precise internal behavior and thereby efficiently operate in real-world nondeterministic environments, the bio-inspired systems also exhibit significant robustness and tolerance against fault, noise, and imprecision sources. As an instance, higher inherent fault tolerance of the bio-inspired artificial neural networks and fuzzy inference engines with respect to hard computing solutions has attracted great attentions again recently[4]. The significant pitfall which is raised during recent years is that despite their inherent tolerance against imprecision, all these bio-inspired and soft algorithms are customarily implemented based on the old-fashioned but very popular hard realization paradigm. For example, there are numerous examples of artificial neural networks and fuzzy engines implemented on either a fully precise hardware or a software running on a deterministic and precise CPU. The significant drawback of this approach is the unnecessary precision guarantee dictated by hard realization that does not improve the overall system performance while it imposes high area, delay, and power overheads (top-right cell of Table I). This shortcoming is exacerbated when the system designer exploits hard realization to implement these soft applications in faulty and nondeterministic nanotechnologies (bottom-right cell of Table I) as will be discussed in the following.

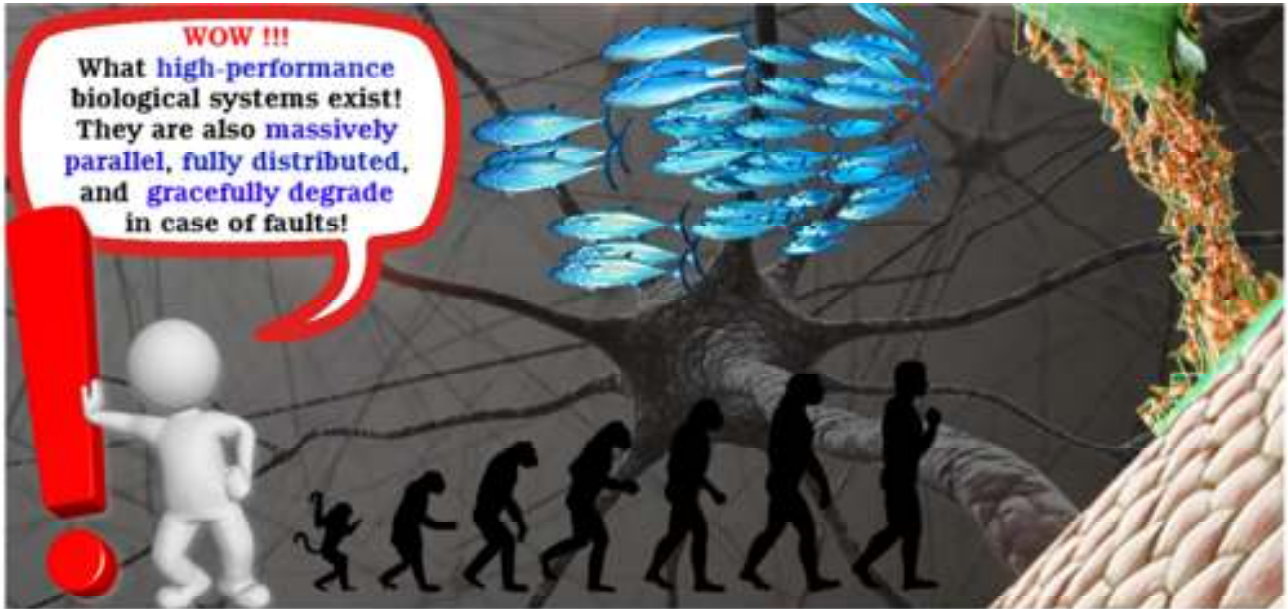

Fig. 1- Valuable features of nature-made biological entities and societies.

## 2-2- Incompatibility of Hard Realization with the Evolving Nanotechnologies

The ever increasing performance demand and complexity of applications is a powerful driving factor behind transistor down-scaling in conventional VLSI technology and also development of alternative implementation technologies such as *Quantum-dot Cellular Automata* and *Carbon Nanotube*. These nanotechnologies provide ultra-high device density (about $10^{12}$ devices/cm$^2$) [5] that facilitates implementing more complex applications with higher performances. They however, also introduce some significant new challenges to the design stack such as higher defect rates and permanent faults (~$10^{-1}$ to $10^{-3}$ defects/part)[2], appearance of transient or intermittent faults, and concurrent occurrence of multiple permanent or transient faults. Downscaled transistors and wires in a nanotechnology reduce the critical charge significantly. The critical charge is the amount of charge necessary to change the state of a node in the circuit. In such a fragile situation, a neutron or alpha particle hit might introduce enough amounts of charge to momentarily toggle a node's state and to cause a transient fault. A transient fault then might be luckily masked after a long enough period, or might unfortunately turn into a soft error in case of latching into a memory element. Historically, soft errors had been a serious concern for only a limited number of applications such as aerospace and defense for a long time; however, it is now a major problem even in commercial products. The significance level of these new challenges can be determined either by noticing the reports that highlight their industrial relevance in existing products [6-8] or more importantly observing practical industrial actions directly relates to soft error mitigation [9-11]. The soft error rate of the future CMOS integrated circuits is predicted to reach 50000 Failures-In-Time (FIT) [12] that is unacceptable for enterprise computing. A FIT is defined as the occurrence of one failure per billion operational hours of a circuit. The importance of the soft errors can also be highlighted by comparing its failure rate with the rate of all other defects and manufacturing failures that range between 1 to 50 FITs [13].

Due to the high failure rate in nanotechnologies, they are not suitable candidates for hard realization of traditional applications (bottom-left cell in Table I) and significantly extra overhead must be paid to achieve acceptable reliability for realizing these applications (Fig. 2). For instance, to implement a traditional hard computing algorithm in a faulty digital VLSI nanotechnology, designers have to simultaneously employ some overpriced fault-tolerant design techniques probably at multiple design levels (i.e. software, architecture, and circuit) to suppress the effects of permanent and transient faults and provide the precision dictated by hard realization. All existing fault tolerant design techniques exploit prevention, detection, and correction methods that result in time, space, and code redundancy in the system. Triple Modular Redundancy (TMR) is a famous mechanism that imposes more than 200% area overheads to detect and correct only single permanent or transient faults. Similarly, a comprehensive accuracy might be achieved only by paying considerable overheads. More importantly, note that the desired accuracy may not be achievable at all in some technologies. Due to the high defect rates and uncontrollable number of transient faults in the emerging nanotechnologies, it is theoretically impossible to determine an upper bound for the number of concurrent faults occurred in an implemented complex system. Therefore, it sounds impossible to preserve full accuracy needed by hard realization even by paying any large amounts of design overheads. Using a nine-modular redundancy scheme as an instance, compels an area overhead of more than 800%, while it increases the system tolerance to handle up to a maximum of four concurrent faults, which is still unacceptable for large systems implemented in nanotechnologies. This issue becomes worse for hard realization of soft application in unreliable nanotechnologies (bottom-right cell in Table I). In such case, the full precision requested by hard realization and achieved by means of expensive fault tolerant design techniques is beyond the necessities of the soft and bio-inspired algorithms!

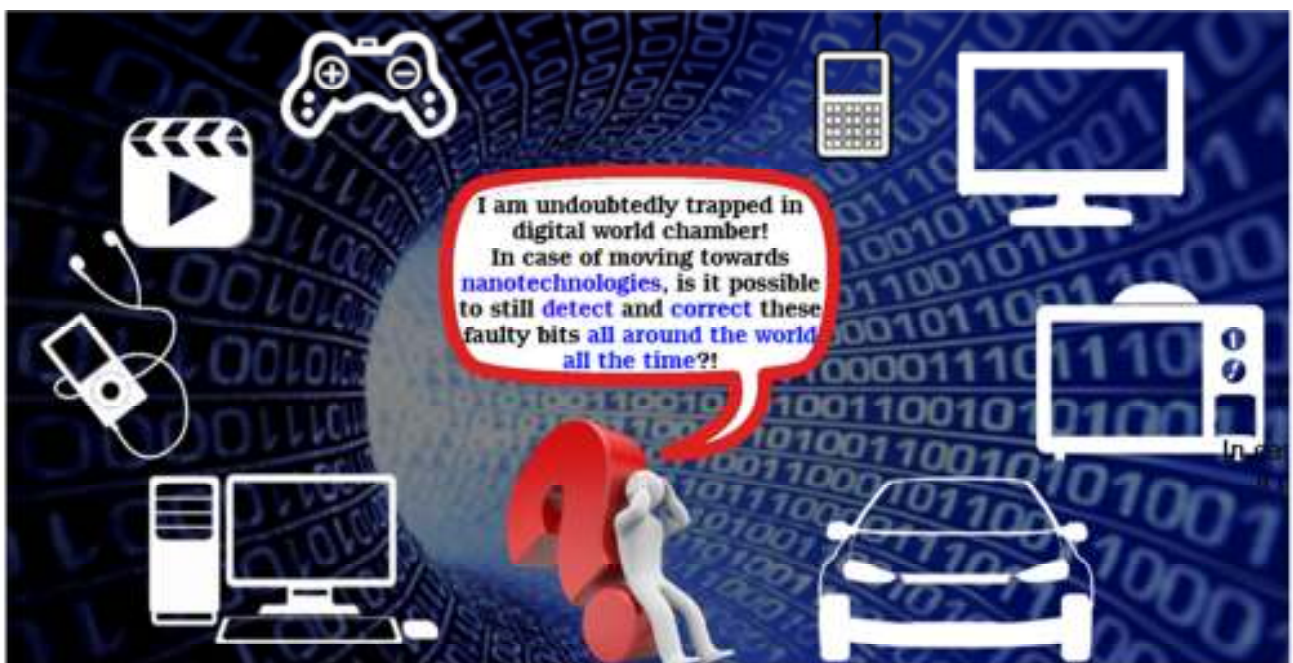

Fig. 2- Inadequacy of the conventional *Hard Realization* paradigm in faulty nanotechnologies.

## 3- "Imprecision Tolerant" Applications: Definition and Preliminary Discussions

Must 2+2 make 4? The answer to this seemingly straightforward question is the key for introducing a new category of applications that we call *Imprecision Tolerant*. Although the answer to this question appears to be a simple "yes", this typical and straightforward answer is not always the best. With a new technical view, the best specific answer to this question depends on the application that uses the result of this arithmetic operation. If this addition is performed to compute an account information in a banking application when updating the current customer account balance (e.g. $2) with the new deposit value (e.g. $2), the answer should be definitely 4. But if it is required in an image processing system to compute the new filtered value of a pixel in an 8-bit gray-scale image based on its current gray level (i.e. 2) and the desired filter value (i.e. 2), the addition result does not have to be 4. Rather, it might be any greater or less value than 4 as long as the human eye could not distinguish its corresponding gray-scale color with the color corresponds to the precise value. For example, it suffices for the addition result to be between 2 to 6, but not 0 or 20 of course.

Based on the same viewpoint which is used to answer above simple question, we have identified a novel category of applications which we call *Imprecision Tolerant*. An application is imprecision tolerant if it tolerates some degrees of imprecision when realized using any implementation bed [3]. This means that the realized system should produce acceptable behavior to the end user despite existing sources of imprecision. Although there are some other previously proposed seemingly similar definitions such as *Fault Tolerant* or *Error Tolerant,* it should be stated that the term imprecision-tolerant is semantically different and defines a broader superset for all those previously defined terms. Technically, all types of faults, manufacturing defects, and environmental damages (which are subjects of the fault and error tolerant contexts) should be treated as some but not all instances of imprecision in a system. The imprecision might be introduced into the system by some other means. As another new imprecision source, we have introduced that the imprecision might be intentionally injected into a system by the designer[16]!

Fig. 3 enumerates some well-known instances of imprecision tolerant applications. As indicated in the figure, the imprecision tolerant concept covers a wide range of significant state-of-the-art research fields in *Artificial Intelligence*, *Digital Signal Processing, Human-Machine Interface*, *Soft Computing*, and so on. To better illustrate the concept of imprecision tolerance, we categorized three different classes of imprecision tolerant applications based on their method that deal with imprecision. The first class contains some applications that can tolerate imprecision up to a threshold and their behavior degrades only if they encounter higher amounts of imprecision. The binary artificial neural network is a good instance of this class. Although the internal behavior of the network degrades with the increase of imprecision, it tolerates imprecision until it has not turned into a functional error by toggling the neuron output. The second class includes those systems whose behavior gracefully degrades as imprecision increases more. In image or voice processing systems as an example, the quality of pixels or voice samples degrades as the imprecision grows inside the system. In such systems, the imprecision is tolerable if the degradation of the pixel colors or the voice samples could not be sensed by human eyes and ears respectively. The third class contains those applications that are capable of absorbing high amounts of momentarily introduced imprecision due to existence of internal or external error recovery feedback mechanisms, while a continuous sequence of small imprecision sources might cause system instability. Although a fuzzy controller might lose its acceptable behavior due to a large pulse of imprecision and issue a wrong command, it can quickly restore its correct behavior again based on input feedbacks which it receives from the system under control.

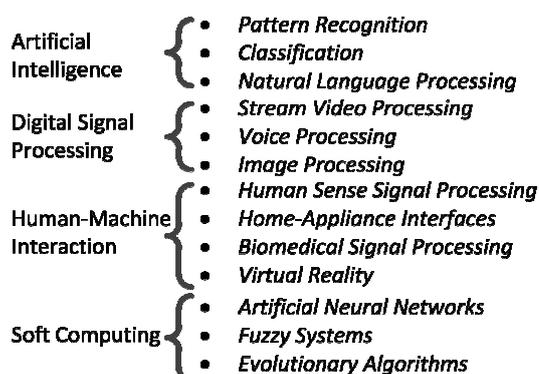

Fig. 3- Some well-known instances of imprecision tolerant applications in different research domains.

**4- Soft Realization: A Bio-inspired Realization paradigm for Imprecision Tolerant Applications**
Investigations on recently published research work show that a new era for soft and bio-inspired algorithms has been started due to their improved performance and less susceptibility against faults and imprecision. Due to increasing importance of these applications that mostly fall in imprecision tolerant category, a new realization paradigm is introduced which is exactly developed based on technical specifications of the emerging applications to cover main shortcomings of hard realization and gain much better performance and cost when implementing such applications in either traditional reliable or emerging unreliable technologies. Due to bio-inspired nature of these algorithms, the best way to find a compatible realization paradigm is to refer to the implementation basics of the same biological computational systems. Although the brain as the most intelligent nature-made computational entity has been the inspiration source of many useful bio-inspirations from different aspects, it is very helpful to observe it once more this time in view of its implementation aspects. The brain is made from huge number of highly connected small computational units called *Neurons*. The collective behavior of this heavily distributed system when dealing with many types of complicated tasks such as recognition, classification, and reasoning, even under extreme ambiguous and uncertain environmental conditions, has tremendous degrees of accuracy, reliability, performance, and real-time response capability. However, this dependable behavior is resulted by running different collective processes on so many imprecise and uncertain biological neurons. The neuron cell as the main building block of the brain consists of a *Body*, some input branches which receive signals from several other neurons called as *Dendrites*, and a single output branch that carries the outgoing electrical signal emitted by the neuron called as *Axon* [14]. The axon of a neuron contacts the dendrites of several other neurons by an electrochemical device called *Synapse* whose task is to transmit electrical signals between two neurons. When electrical signal reaches the synaptic point, it releases some chemical substances known as *Neurotransmitters* which open molecular gates on the dendrites of the receiving neuron and causes electrically charged particles or ions to flow in. The received ions from different synapses cause a voltage difference between neuron body and external environment. The neuron fires and propagates an electrical signal called as *Pulse* or *Spike* along its axon when this voltage difference is larger than a defined threshold. This fundamental cycle lasts between 3 to 50 ms depending on the type of ions [14]. The neurons can fire up to 500 times/sec in normal operation while the firing rate decreases to approximately 1 to 5 times/sec when there is no incoming signal [14]. The neural connections should also be considered as the brain distributed memory. The communication structure of the brain is also based on transferring chemical ions and weak electrical signals in orders of tens of millivolts.

This brief biological structural description implies that when dealing with many types of complex but soft and imprecise tasks, the tremendous degrees of precision, performance, and reliability provided by the brain are the collective result of a massively parallel network of many low precision, slow, and uncertain processing units, memories, and connections. The promising note which encourages us to follow this natural computation paradigm is that it is also supported by many other biological phenomena. For example, there are many instances of biological societies with seemingly intelligent and complex global behaviors while they are basically composed of many unreliable and much simpler individuals such as ants or fishes. The very important inspiration from brain implementation aspects is that regardless of the implementation technology, it is not mandatory to exploit completely deterministic and precise building blocks and implementation environment to achieve high amounts of performance and reliability. It is also very important to note that this is true only for soft and bio-inspired or in more technical way imprecision tolerant tasks. Similar to brain, this inspiration will not work for the hard and rigorous algorithms and applications with rigid and strict accuracy requirements. Based on our daily experiences, although the raw brain of a one year old baby is capable of simultaneously real-time execution of many complicated but imprecise tasks (such as recognition and data fusion of the noisy face and voice of the parents), it is a time consuming and boring task even for the mature and experienced brain of a scientist to perform even a very simple task with high precision requirements, such as the addition of two 3-digit numbers!

Unlike hard realization that relies on full precision and determinism, a brain inspired approach may relax the design and implementation process to decrease complexity and increase performance. This novel realization paradigm named as *Soft Realization* due to its similar foundations with famous *Soft Computing* methodologies. The main difference between soft computing and soft realization is that the former contributes in algorithmic level while the latter covers implementation level issues. *Soft Realization* is applied to an emergent collection of design and implementation methodologies or techniques which trade the inherent tolerance of an imprecision tolerant application to simplify its implementation process or decrease its implementation costs regardless of the implementation technology, without degrading its performance to an undesirable region from the end user viewpoint.

At the first glance the elasticity of soft realization seems to be unusual in comparison with long-standing, rigid, and precise conventional hard realization. However, its main advantage relates to its broad compatibility with emerging applications and nanotechnologies, where exploiting hard realization compels high amounts of overheads as discussed before (summarized in top-right and bottom-right cells of Table I). From the first projections of the soft realization and the imprecision tolerant concepts as possibly feasible research fields [15] up to now, we have introduced two basic well-defined soft realization methodologies for efficient realization of the imprecision tolerant applications on both precise/deterministic [16] and imprecise/nondeterministic implementation technology [3]. Fortunately, both methods are appreciated and supported by many other active VLSI design researchers during the next years. Our achieved experimental results as well as findings of many other following interested researchers [23-73] prove the superiority of different soft realized imprecision tolerant applications with respect to their corresponding hard realized rivals with little or no performance degradation.

**5- Pioneer Instances of the Proposed Soft Realization Methodologies**

This section surveys the advantages of the two significant introduced instances of cutting edge soft realization methodologies [3,16]. They demonstrate not only feasibility but also effectiveness of the soft realization to overcome the shortcomings of the implementation of soft applications in both reliable and unreliable technologies (summarized in top-right and bottom-right cells of the Table I, respectively). The proficiency of both methodologies is verified by realizing some crucial and computation-intensive instances of imprecision tolerant applications, a face-recognition artificial neural network, and the defuzzification block of a hardware fuzzy processing engine [16]. The face-recognition test case contains a large scale three-layer neural network. It consists of nine output neurons, eight internal-layer neurons each with 960 inputs that correspond to different pixels of a 32 by 30 gray-scale photo. The next test case is a hardware defuzzification block as the most demanding part of a *Mamdani* fuzzy processing engines that defuzzifies the fuzzy input value based on *Weighted Plateau Average* algorithm [17].

**5-1- A Soft Realization Methodology in Traditional VLSI technology: Bio-inspired Imprecise Computational Blocks**

Despite inherent incompatibility of the hard realization with emerging applications (as indicated in top-right cell in Table I), exploiting soft realization for implementing imprecision tolerant applications in traditional digital VLSI technology simultaneously improves both system cost and performance remarkably. As a very plausible instance in digital VLSI territory, the soft realization modifies the conventional ternary "area-delay-power" trade-off to a new four-way "area-delay-power-imprecision" trade-off when implementing an imprecision tolerant application. Based on the new dimension added to the conventional trade-off, the designer can intentionally introduce or inject the imprecision in any desired implementation process step up to a tolerable level to achieve many other benefits such as less number of devices, faster circuits, less power consumption, or any other probable design criteria. Based on this useful hint, we have introduced some imprecise VLSI computational blocks (e.g., adders, multipliers) named as *Bio-inspired Imprecise Computational* (BIC) blocks [16]. The internal structure of a BIC is intentionally modified to provide only an approximation of the result instead of its precise value. This results in much better physical properties in terms of area, delay and power with respect to precise computational blocks. The BICs should be armed with one or more imprecision parameter(s) [22] which can be used to fine-tune the amount of imprecision introduced by the BIC based on maximum tolerable imprecision level of each imprecision tolerant application. Increasing the BIC imprecision parameter(s) result in higher imprecision as well as better physical properties of the component [18, 19]. In a similar manner, figure 4 demonstrates some other probable soft realization techniques which are developed by reducing the precision at some other design and implementation steps of the traditional digital hardware and software design flows. Figure 4-a shows that it is possible to propose some instances of imprecise hardware description languages, synthesis tools and low-level building blocks. As an instance, an imprecise synthesis tool might either directly exploit imprecise components as its library components or exploit some simplified and faster versions of the mostly time-consuming NP-complete synthesis algorithms to perform an approximate or soft synthesis instead of time-consuming conventional hard synthesis. In software implementation as another important instance (figure 4-b), it is possible to develop some imprecise programming languages, compilers, and central processing units.

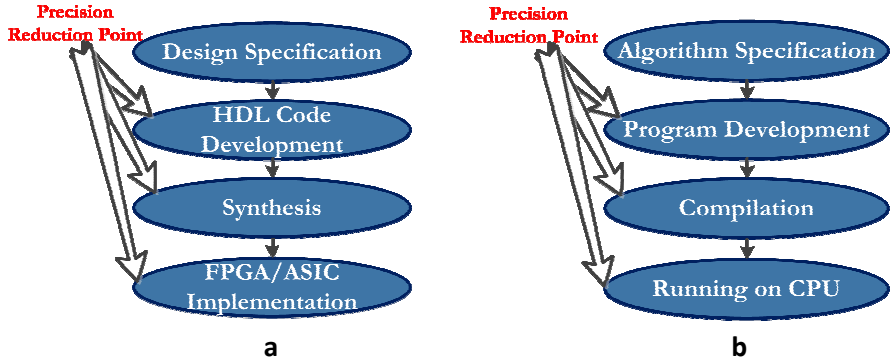

Fig. 4- Some other probable precision reduction (imprecision injection) points in (a) digital hardware design flow, (b) software design flow

Today, proposing new BICs or novel techniques to efficiently exploit them to implement different applications have turned into an active and popular research area. A categorized survey of some recently published papers (since 2015) is included in supplementary information[23-73] to emphasize the significance and broadness of this new topic. Although some other names such as *Approximate* [20] or *Sloppy* [21] are later proposed for this type of blocks, the term *Imprecise* is still preferred because it covers much broader area with respect to approximate or sloppy. As an instance, the approximate computing can be only viewed as a subclass of the imprecise computing in which, the designer intentionally and predictably changes the precise computational circuits in a deterministic manner to provide a result approximation instead of its precise value. However, in the next section we introduce another more important branch of the imprecise computing while the computation results are acceptable despite probably existing multiple unrecovered soft errors which cannot be referred to as approximation anymore. Besides, the term imprecision also demonstrates the tight correspondence between BICs and imprecision-tolerant applications in a stronger manner. To provide a brief illustration about efficiency of exploiting the BICs, Table II presents the improvements in physical properties of the hardware implementation of the face recognition neural network and the defuzzification block when exploiting BICs instead of precise computational blocks [16]. The table results show that the precise neural network is about 30% larger, 54% slower, and has about 101% worse area×delay product with respect to the BIC neural network. Also the precise defuzzification block has about 108%, 54%, and 220% larger area, delay and area-delay product respectively with respect to its precise rival. It is noteworthy to emphasize these considerable improvements are achieved without any significant performance degradation in both applications [16]. More details about the achieved improvements are included in the supplementary information for more convenience.

TABLE II- Improvements in physical properties of the hardware implementation of two critical imprecision tolerant applications using BICs and conventional precise components.

| Application | *Physical Property Improvement (%)* | | |
|---|---|---|---|
| | *Area* | *Delay* | *Area × Delay* |
| **Neural Network** | 54 | 30 | 101 |
| **Defuzzification** | 108 | 54 | 220 |

## 5-2- A Soft Realization Methodology for Emerging Nano-level VLSI Technologies: Relaxed Fault-Tolerant

This section demonstrates another important instance of the soft realization paradigm. Due to the high cost overheads of traditional fault-tolerant techniques in hard realization, a new methodology is specifically developed to reduce fault tolerant costs of imprecision tolerant applications in faulty environments such as VLSI nanotechnologies. According to this novel soft realization methodology, called *Relaxed Fault-Tolerant* (RFT)[3], it is unnecessary to correct or even detect all faults occurred in a system. Moreover, it is not necessary to recover from a fault if it doesn't have significant impact on the quality of system behavior or if the correct behavior of the system may be restored in a timely manner[3]. Although much attention has been recently paid to the BICs by researchers, we believe that the RFT concept plays a significant role as we mandatorily move towards deeper nanotechnologies, where reliable hard realization becomes more and more impractical (bottom-left and bottom-right cells of Table I). It might be interesting to investigate the similarity of the underlying ideas of the BIC and RFT as two different instances of the imprecise computing. Based on soft realization paradigm, the BIC proposes to intentionally inject some imprecision sources into design components to reduce the total system implementation cost, while the RFT improves the fault tolerant design costs by preventing to correct and even detect some unintentional existing imprecision sources (e.g., transient or permanent faults).

RFT suggests exploiting the relaxed versions of many existing fault-tolerant design techniques instead of the original precise versions, while optimizing cost and efficiency of imprecision tolerant applications in faulty technologies. In an example computational block, a straightforward relaxation method is to more focus on the soft errors affect more important bits such as sign bit or some most significant bits of the result[3]. This leads to some RFT computational blocks that exploit modified fault tolerant techniques with different fault correction levels to compute different result bits of a single computation. Table III demonstrates area, delay, and area×delay improvements achieved for the *Relaxed Triple Modular Redundancy* hardware in face recognition neural network and the defuzzification block with respect to full *Triple Modular Redundancy*[3]. The results indicate that the neural network TMR model is about 155% larger and 50% slower; moreover, it results in a total of 288% worse area-delay-product as compared to the RTMR. Similarly, the TMR model of a defuzzification application has about 89%, 39%, and 163% larger area, delay, and area-delay-product overheads than those of the RTMR model. These significant improvements are achieved at the cost of negligible performance degradations. More details about the achieved improvements are included in the supplementary information.

Table III- Improvements in physical properties of the hardware implementation of two critical imprecision tolerant applications using RTMR and TMR fault tolerant design techniques.

| Application | Physical Property Improvement (%) | | |
|---|---|---|---|
| | Area | Delay | Area × Delay |
| Neural Network | 155 | 52 | 288 |
| Defuzzification | 89 | 39 | 163 |

## 6- Discussion

In this paper, we first clarified the technical specification of a broad category of mostly emerging, soft and bio-inspired applications named as *Imprecision Tolerant* applications. The main common feature of this new category which includes many applications from a diverse range of different research fields is that they can provide acceptable behavior to the end-user despite probable existing imprecision sources and nondeterministic effects when they are realized by any implementation technology. The main underlying concepts of a newly proposed realization paradigm named as *Soft Realization* are then discussed in the paper. The soft realization is mainly developed for efficient implementation of imprecision tolerant applications. The golden rule of the soft realization is that it changes the conventional area-delay-power trade-off that pervasively exists in traditional implementation strategy to an area-delay-power-precision four-way trade off. In other words and in opposite of the traditional realization paradigm, it tries to activate and exploit the inherent tolerance of the imprecision tolerant applications against imprecision and nondeterminism to trade it and achieve considerable realization improvements in terms of system physical costs. Although the soft realization is a technology independent implementation paradigm, two pioneer instances of its application in digital implementation as the most popular implementation technology is presented in this paper to demonstrate its applicability as well as efficiency. The implementation results of a face recognition artificial neural network and a defuzzification block of a hardware fuzzy processor as the two computation-intensive instances of the imprecision tolerant applications on a traditional reliable VLSI technology as well as an unreliable nanotechnology are included in the paper. Our presented experimental results as well as the results of many other following researchers prove that exploiting the soft realization leads to considerable physical improvements of the imprecision tolerant applications in terms of area, delay and power consumption with negligible performance degradation.

**Supplementary Information** is available in the online version of the paper.

**Acknowledgements** We would like to thank profeesor Caro Lucas (RIP) for his helpful discussions and advice, as well as his kind and great support during establishmnet of the Soft realization idea foundations.

**Author Contributions** H. R. M. and S. M. F. (RIP) developed the main Soft Realization idea. H. R. M. introduced and developed the BIC and RTMR models of the face-recognition neural network and defuzzificaion blocks and wrote the paper. S. M. F. advised the project and revised the paper. M. N. B. revised the paper and proposed some improvements.

## A Soft Realization Methodology in Traditional VLSI technology: Bio-inspired Imprecise Computational Blocks [16]

Recently, a great attention is paid to development of novel BIC structures such as adders [23-38], multipliers [39-53], and dividers [54-56], as well as proposing CAD tools as software frameworks for better exploitation of BICs [57-66]. Also there are many useful survey papers which categorize and demonstrate different BIC related research areas [67-73]. To demonstrate feasibility as well as efficiency of this interesting soft realization methodology, we have proposed an imprecise adder called *Lower-part OR Adder* (LOA), and an imprecise multiplier called *Broken-Array Multiplier* (BAM) [16]. The LOA divides a p-bit addition into two m-bit and n-bit smaller parts (m+n=p). As shown in Fig. 4 a p-bit LOA exploits a regular precise adder to compute precise values of the 'm' most significant bits of the result along with some OR gates that approximate the 'n' least significant result bits (lower-part). For a LOA with a constant *Word-Length* (WL or 'p'), it is obvious that increasing 'n' or *Lower-Part Length* (LPL) imprecision parameter decreases the physical properties (area, delay, and power consumption) of the LOA while at the same time increases the imprecision. Fig. 5 demonstrates the basic structure of a 6×6 BAM which is constructed based on structure of the famous *Array Multiplier*. An array multiplier consists of m×n array of similar *Carry-Save Adder* (CSA) cells while the BAM breaks the CSA array and omits some cells that lead to a smaller and faster circuit with approximate results. As shown in Fig. 5, the number and position of the omitted hatched cells depend on two imprecision parameters: *Horizontal Break Level* (HBL) and *Vertical Break Level* (VBL). The HBL=0/VBL=0 means that there is no horizontally/vertically omitted CSA cell. Increasing the HBL/VBL moves the horizontal/vertical dashed line downward/leftward and all cells which fall above/right-side of the dashed line are omitted.

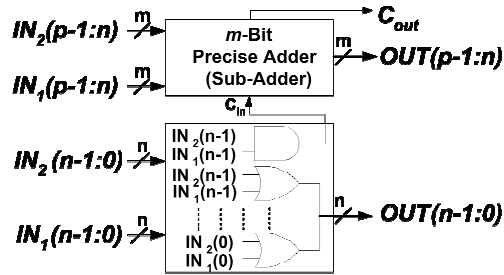

Fig. 4- Hardware structure of the Lower-part-OR Adder (LOA) [16].

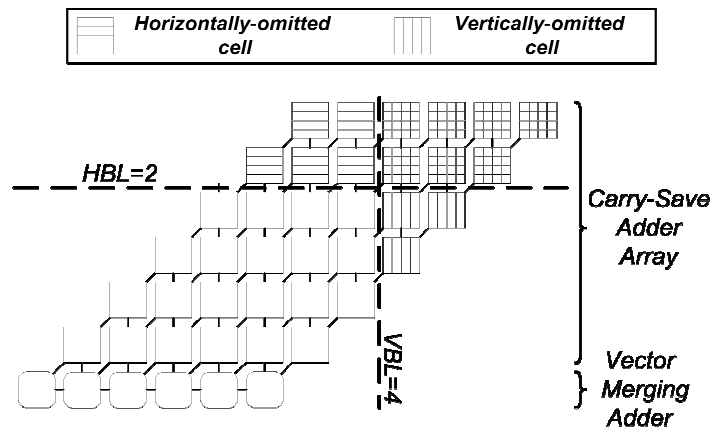

Fig. 5- Hardware structure of the Broken-Array Multiplier (BAM) [16].

## BIC Implementation of the Face-Recognition Neural Network

Table IV includes the error simulation and synthesis results of the precise and BIC models of the face-recognition neural network with WL=9 [16]. The results of the BIC model are presented for some of its best acceptable combinations of the imprecision parameters (LPL, HBL, and VBL) in which the BIC model works properly. The first column determines the type of the model where 'P' stands for precise and 'B' for BIC models. The next four columns include the parameter values of the models. The 6$^{th}$ and 7$^{th}$ table columns include the Mean Squared Error (MSE) and Training Epoch number (TE#) simulation results of the precise and BIC models. The three last columns of table include the data-path area (without registers), critical-path delay, and area-delay product of the models as they are synthesized on 0.13μ CMOS technology library cells using Leonardo Spectrum synthesis tool. The table results show that BIC models can nearly provide the same MSE and TE# with respect to the precise model. The interesting note is that although the BIC models provide similar performances based on these error metrics, they have much better physical properties with respect to the precise model. Based on table values, the normal implementation of the neural network is always faster and has respectively 12%, 20%, and 30% higher clock periods than the three BIC implementations while it requires larger area. It is also clear that the precise model has 54%, 30%, and 101% higher area, delay, and area-delay product with respect to the BIC#1 while both recognize the input faces properly and have similar internal behaviors. For more convenience, Fig. 6-a illustrates the size of precise and imprecise sections of the LOA and BAM in B#1 (LPL=2, HBL=2, and VBL=6). In each block, the imprecise sections are marked in black. As another instance, the precise model has about 20% longer critical path delay, 43% more gate count, and 72% worse area-delay product with respect to the BIC#2 as another suitable parameter combination of the BIC model.

TABLE IV- Error values and physical properties of the face-recognition NN precise and BIC models.

| NN Model Type | Parameter Values | | | | NN Simulation Results | | NN Synthesis Results | | |
|---|---|---|---|---|---|---|---|---|---|
| | WL | LPL | HBL | VBL | MSE | TE# | Area (# of gates) | Delay (ns) | Area × Delay |
| P#1 | 9 | - | - | - | 0.0008 | 68 | 4000 | 17.98 | 71920 |
| B#1 | 9 | 2 | 2 | 6 | 0.0008 | 49 | 2586 | 13.84 | 35790 |
| B#2 | 9 | 3 | 1 | 6 | 0.0006 | 69 | 2783 | 14.96 | 41633 |
| B#3 | 9 | 4 | 0 | 5 | 0.0006 | 88 | 3179 | 16.06 | 51054 |

**BIC Implementation of the WPA Defuzzification Block**

Table V includes the error simulation and synthesis results of the WPA precise and BIC models in some optimum points of the parameter space when synthesized with 0.13μ CMOS library cells using Leonardo Spectrum synthesis tool [16]. The first column determines the type of the model ('P' for precise and 'B' for BIC models) while the next 4 columns include the parameter values. The 6$^{th}$ to 8$^{th}$ columns include the Mean Absolute Error (MAE), Absolute Error Variance (AEV), and MAXimum absolute error (MAX) simulation results of the defuzzified output value with respect to the whole fuzzy number width (FuzzyWidth) in percentage. The last three columns of the table show area, delay, and area×delay product of each model respectively. The results show that all imprecise models provide much better clock frequencies and implementation areas with respect to the precise model of the same WL while have slightly lower accuracy. As an example, the BIC#1 and BIC#3 instances provide less than 0.8% and 1.1% worse average and variance errors with respect to the precise model of the same WL (P#1) while having 35% and 54% higher clock frequencies. Also the hardware implementation of the precise model has an area×delay product of about 130% and 220% higher than BIC#1 and BIC#3 models respectively. Fig. 11-b illustrates the size of precise and imprecise sections of the LOA nd BAM for B#3 (LPL=2, HBL=1, and VBL=6). Another interesting observation is that the P#1 provides 37% and 68% higher area×delay product with respect to two BIC models with higher WL (B#5 and B#6) while their accuracy factors are very similar.

Table V- Error values and physical properties of the WPA defuzzification block precise and BIC models.

| WPA Model Type | Parameter Values | | | | WPA Simulation Results | | | WPA Block Synthesis Results | | |
|---|---|---|---|---|---|---|---|---|---|---|
| | WL | LPL | HBL | VBL | MAE | AEV | MAX | Area (# of gates) | Delay (ns) | Area × Delay |
| P#1 | 6 | - | - | - | 2.43 | 2.80 | 15.9 | 1686 | 10.16 | 17129 |
| P#2 | 7 | - | - | - | 2.45 | 3.04 | 14.2 | 2328 | 12.59 | 29309 |
| B#1 | 6 | 2 | 1 | 5 | 3.05 | 3.64 | 20.3 | 994 | 7.48 | 7435 |
| B#2 | 6 | 2 | 2 | 5 | 3.44 | 4.15 | 20.3 | 887 | 6.38 | 5659 |
| B#3 | 6 | 2 | 1 | 6 | 3.29 | 3.91 | 20.3 | 811 | 6.59 | 5344 |
| B#4 | 6 | 2 | 2 | 6 | 3.66 | 4.31 | 20.3 | 757 | 5.92 | 4481 |
| B#5 | 7 | 2 | 1 | 6 | 2.54 | 3.01 | 15.1 | 1358 | 9.21 | 12507 |
| B#6 | 7 | 2 | 2 | 6 | 2.53 | 2.84 | 15.2 | 1252 | 8.12 | 10166 |

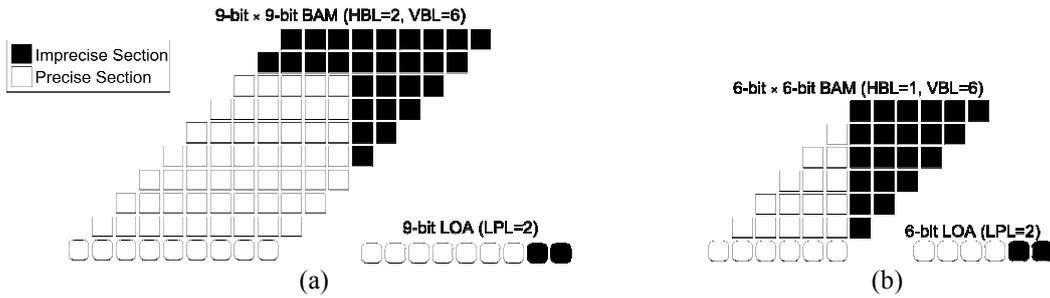

Fig. 6- Precise and Imprecise portions of the exploited BAM and LOA in BIC models of two IT applications: a) Neural Network, b) Defuzification [9].

**A Relaxed Fault Tolerant Design Technique: Relaxed Triple Modular Redundancy [3]**
To show the applicability and performance of the RFT as a significant instance of soft realization in faulty nanotechnologies, we developed a relaxed version of the *Triple Modular Redundancy* (TMR) as a well-known fault tolerant design technique. The TMR technique simultaneously exploits three similar modules, applies similar patterns to their inputs, and then votes between their outputs to achieve the correct output result even if one of the modules fails. Instead of applying the TMR technique to all parts of a computational component, the *Relaxed Triple Modular Redundancy* (RTMR) suggests to protect only some of the most significant bits of the computational block using TMR while the remaining least significant bits are left unprotected. Although such a block might observe many errors on its least significant bits due to lack of protection against soft errors, the experimental results show that it can be used to implement imprecision tolerant applications without losing considerable amounts of performance. Just similar to BIC imprecision parameters, any RTMR computational block has one or more parameters that determine the boundaries of the protected and unprotected portions of the block. Increasing these parameters extend the unprotected area while at the same time improves the FT implementation cost of the block. The values of these parameters for each RTMR block should be determined based on the imprecision-tolerance level of the application which is determined through system-level simulations.

**General Illustration of a Relaxed Fault Tolerant Adder**
Fig. 7 illustrates the general structure of a q+m+n bit RFT adder. It exploits three different smaller sub-adders with different fault-tolerance levels to compute different result bits of a single addition. Different fault-tolerance levels might be translated to using different fault tolerant design techniques and paying different amounts of redundancy (in time, space, code) for implementing each smaller sub-adder. The relaxation process might be applied to many of the existing fault tolerant design techniques at different levels such as device, circuit, and system levels. For example, a device level RFT technique proposes to use transistors with different gate sizes for different sub-adders of the whole RFT adder (i.e. larger transistors for more significant bits), a circuit level RFT technique implies using different self-checking and self-correcting circuits for different sub-adders. Also according to existing system-level fault tolerant design techniques, a system-level RFT might be defined as using different numbers of redundant blocks for each sub-adder.

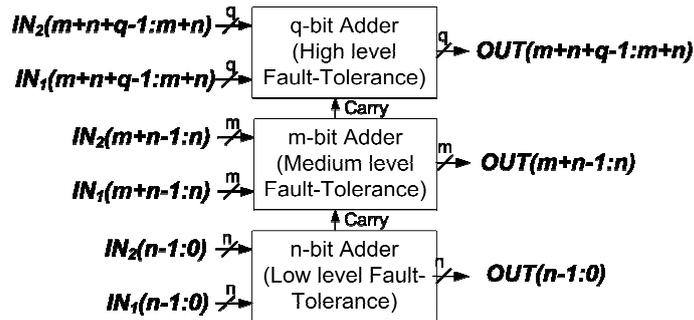

Fig. 7- Structure of a RFT adder with three different fault-tolerance levels [3].

**Relaxed Triple Modular Redundancy (RTMR) Adder and Multiplier Models**
Fig. 8 illustrates the main structure of a *Relaxed-TMR Ripple Carry Adder* (RRCA) which has similar basics with respect to LOA. It is basically a ripple carry adder while it has two different fault-tolerance levels. As shown in the figure, the least significant bits of the result are computed without any protection technique (fault-probable area). The reason is that any change in these bits that might occur due to soft errors, does not significantly affect the overall result and the produced error does not destroy correct functionality of the whole imprecision tolerant application. On the other hand, the most significant bits of the result are protected using TMR scheme (fault-protected area), which highly prevent soft errors to change these result bits. The *Adder Unprotected-Length* (AUL) parameter determines the boundary between fault-probable and fault-protected sections. As AUL increases, the length of the fault-probable and fault-protected sections increases and decreases respectively. The important note is that changing AUL also highly affects the RRCA physical properties.

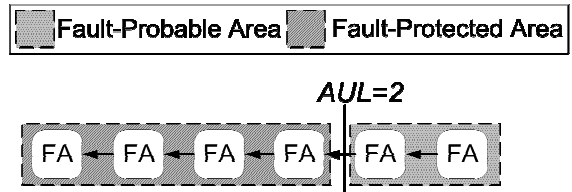
Fig. 8- Relaxed-TMR Ripple Carry Adder (RRCA) structure [3].

The *Relaxed-TMR Array Multiplier* (RAM) is an instance of the RFT multipliers with similar basics of BAM. The structure of a 5×5 RAM that is similar to the structure of an array multiplier is shown in Fig. 9. A RAM exploits two different fault-tolerant levels to protect different CSA cells. As shown in the figure, the number and position of the fault-probable CSA cells depend on two parameters: *Horizontal Unprotected Length* (HUL) and *Vertical Unprotected Length* (VUL). The HUL=0 means that there is no horizontally fault-probable CSA cell. Increasing the HUL moves the horizontal dashed line downward while all cells that fall above the dashed line are not fault protected. Similarly VUL=0 implies that there are no vertical fault-probable CSA cells. As VUL increases, the vertical dashed line moves left while all separated right-side cells are not fault-protected. A similar technique might be applied to other multipliers such as *Wallace-tree* and *Booth-encoding* to build new instances of RFT multipliers [12].

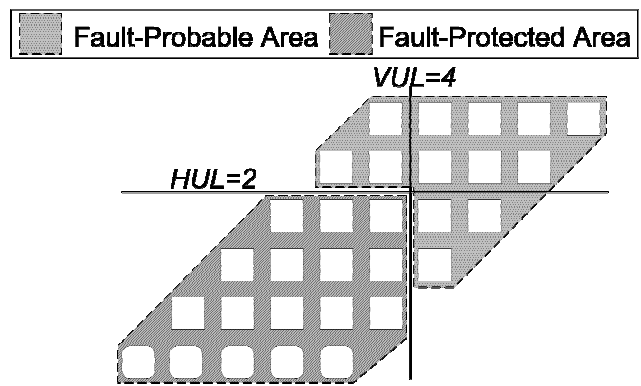
Fig. 9- Relaxed-TMR Array Multiplier (RAM) structure [3].

**RTMR Implementation of the Face-Recognition Neural Network**
Figure 10 illustrates the MAE and correct classified patterns of the RTMR model for a wide range of parameter values when the probability of occurring a soft error in each circuit node (*Perr*) is $10^{-5}$ [3]. The dashed lines on the figure are drawn to demonstrate general trend of changes. The results indicate that increasing AUL by one unit significantly degrades the network output quality that results in step changes. It is also indicated that for a constant AUL, increasing HUL destroys the output quality with smaller rates than AUL. And finally the figure shows that for a constant AUL and HUL, increasing VUL degrades the output quality much lower than AUL and HUL.

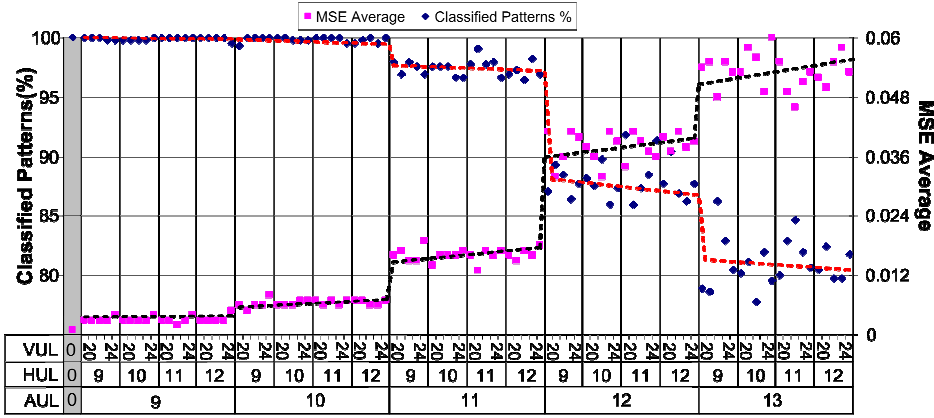

Fig. 10- Correct classification percent and average MSE of RTMR model for training set when $Perr=10^{-5}$ [3].

Table VI includes the simulation and synthesis results of the face recognition neural network which is used in TMR and RTMR realizations when $Perr=10^{-5}$ [3]. The first column determines the model type and the next four columns include parameter values of the model. Columns 6 and 7 include the simulation results and the last three columns include the synthesis results of a single hardware neuron [3]. The simulation results show that the RTMR model has the same behavior as that of the TMR model while at the same time, the synthesis results indicate that the implementation costs of RTMR model is significantly reduced with respect to TMR model due to using a RFT technique instead of a traditional FT technique. The last three column results show that TMR model has 155% larger datapath, 52% longer critical path, and 288% larger area-delay-product with respect to RTMR model. The table results clearly demonstrate that RTMR model has an acceptable behavior and also a good fault tolerance level with respect to TMR model while is implemented with a much smaller hardware and operates with double speed. To better illustrate the achieved physical properties improvement in RTMR model, Fig. 11-a illustrates the size of the protected and unprotected portions of the exploited RRCA and RAM when AUL=5, HUL=8, and VUL=18. In each block, the unprotected portions are marked in black.

TABLE VI- Simulation and physical properties of the NN TMR and RTMR models for $Perr=10^{-5}$.

| NN Model Type | Parameter Values | | | | NN Simulation Results | | NN Synthesis Results (A single hardware neuron) | | |
|---|---|---|---|---|---|---|---|---|---|
| | AUL | HUL | VUL | Perr | Training Set Average MSE | Training Set Correct Classification Percent | Gate-Count | Gate-Delay | Area × Delay |
| TMR | 0 | 0 | 0 | $10^{-5}$ | 0.001 | 100 | 6552 | 201 | 1316952 |
| RTMR | 5 | 8 | 18 | $10^{-5}$ | 0.001 | 100 | 2568 | 132 | 338976 |

**RTMR Implementation of the WPA Defuzzification Block**

Table VII includes the simulation and synthesis results of the defuzzification block consists of a *Multiply-Accumulate* unit which is used in TMR and RTMR realizations when $Perr=10^{-4}$. Columns 6 to 8 include the simulation results and the last three columns include the synthesis results. The simulation results show that the defuzzification TMR model has the same behavior as that of RTMR model while proposes 89%, 39%, and 163% worse area, delay, and area-delay product respectively. In a similar manner, Fig. 11-b illustrates the size of the protected and unprotected portions of the exploited RRCA and RAM when AUL=4, HUL=4, and VUL=5.

TABLE VII- Simulation and physical properties of the Defuzzification block TMR and RTMR models for Perr=$10^{-4}$.

| Model Type | Parameter Values | | | | Defuzzification Block Simulation Results | | | Defuzzification Block Synthesis Results | | |
|---|---|---|---|---|---|---|---|---|---|---|
| | AUL | HUL | VUL | Perr | MAE | AEV | MAX | Gate Count | Gate Delay | Area × Delay |
| TMR | 0 | 0 | 0 | $10^{-4}$ | 2.46 | 2.75 | 32 | 1782 | 96 | 171072 |
| RTMR | 4 | 4 | 5 | $10^{-4}$ | 2.49 | 2.85 | 48 | 940 | 69 | 64860 |

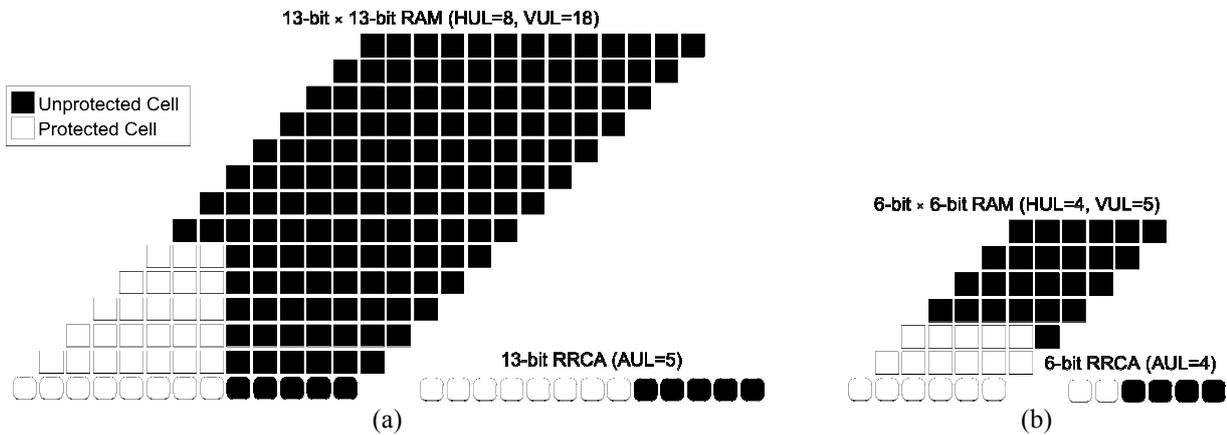

Fig. 11- Protected and unprotected portions of the exploited RAM and RRCA in RTMR models of two IT applications: a) Neural Network, b) Defuzzification.